\documentclass[]{eptcs}
\providecommand{\event}{SCSS 2012} 
\usepackage{breakurl}             
\usepackage{amsmath}
\usepackage{amsfonts}
\usepackage{graphicx}
\usepackage{color,soul}
\usepackage{url}
\usepackage{hyperref}
\usepackage{appendix}

\title{Formal Probabilistic Analysis of a Wireless Sensor Network for Forest Fire Detection}
\author{Maissa Elleuch
\institute{National School of Engineers of Sfax\\
Sfax University\\
Sfax, Tunisia}
\email{maissa.elleuch@ceslab.org}
\and
Osman Hasan
\institute{Dept. of Electrical \& Computer Engineering\\
Concordia University\\
Montreal, Quebec, Canada}
\email{o\_hasan@ece.concordia.ca}
\and
Sofi\`{e}ne Tahar
\institute{Dept. of Electrical \& Computer Engineering\\
Concordia University\\
Montreal, Quebec, Canada}
\email{tahar@ece.concordia.ca}
\and
Mohamed Abid
\institute{National School of Engineers of Sfax\\
Sfax University\\
Sfax, Tunisia}
\email{mohamed.abid@enis.rnu.tn}
}
\def\titlerunning{Formal Probabilistic Analysis of a Wireless Sensor Network for Forest Fire Detection}
\def\authorrunning{Maissa Elleuch, Osman Hasan, Sofi\`{e}ne Tahar \& Mohamed Abid}
\begin{document}
\maketitle

\begin{abstract}
Wireless Sensor Networks (WSNs) have been widely explored for forest fire
detection, which is considered a fatal threat throughout the world. Energy
conservation of sensor nodes is one of the biggest challenges in this
context and random scheduling is frequently applied to overcome that. The
performance analysis of these random scheduling approaches is traditionally done by
paper-and-pencil proof methods or simulation. These traditional techniques
cannot ascertain 100\% accuracy, and thus are not suitable for analyzing a
safety-critical application like forest fire detection using WSNs.
In this paper, we propose to overcome this limitation by applying formal probabilistic
analysis using theorem proving to verify scheduling performance of
a real-world WSN for forest fire detection using a k-set randomized algorithm
as an energy saving mechanism.
In particular, we formally verify the expected values of coverage intensity, the upper
bound on the total number of disjoint subsets, for a given coverage intensity,
and the lower bound on the total number of nodes.
\end{abstract}

\section{Introduction}
Forest fires are considered to be one of the worst natural disasters throughout the world.
They threaten forests, animals, and people, and cause a lot of environmental degradations.
According to recent statistics \cite{Turk_center}, more than 100,000 wildfires occur annually.
For example, in Tunisia, 103 fires destroyed 287 hectares of forests just between May 1, 2012 and July 25, 2012 \cite{Tunis_center}.

For early detection of such fires and thus their prevention, we require robust communication mechanisms that meet critical real-time constraints \cite{fire_center}.
While classical satellite-based monitoring technology \cite{Li_00} has already shown its limitations, Wireless Sensor Networks (WSNs) have emerged as a promising alternative.
A typical wireless sensor network \cite{Yick_08} is composed of small, battery-powered devices, called sensors, which are wirelessly connected over the field of interest (here the forest).
Sensors take measurements of the environment such as temperature and humidity, and communicate altogether in order to report the collected data to a remote user, where appropriate decisions can be made.
Thereby, the outbreaks of a fire can be detected as well as predicted with a small delay using WSNs.
The sensor nodes are usually randomly deployed with a high density in the forest
to ensure full coverage.
Such heavy deployment results in additional energy losses since it is highly probable that the same region would be covered simultaneously by many active sensor nodes. Random scheduling \cite{Jain_07} has been proposed for energy conservation while maintaining WSN coverage. The main idea behind the random scheduling is to split the whole WSN into $k$ sub-networks of nodes that work alternatively.

The k-set randomized scheduling algorithm has been already proposed for use in forest fire detection applications using Wireless Sensor Networks \cite{forest_app,Xiao_09}.
The random feature of the k-set randomized scheduling makes it very challenging
to analyze for all possible cases. Hence, paper-and-pencil based
probabilistic techniques have been traditionally used.
In such analysis, a mathematical model
is built by first identifying the required random variables and the corresponding performance
attributes, then, a rigourous analysis based on the theoretical foundations of probability is done.
Simulation, using the Monte Carlo method \cite{Mackay_98}, is finally used to validate the analytical results.
This widely used validation method is based on approximately answering a query on a probability distribution by analyzing repetitively a large number of samples.
Statistical quantities, such as
expectation and variance, may then be calculated, based on the data collected during
the sampling process, using their mathematical relations in a computer.
Due to the inherent inaccuracies of simulation coupled with the rounding errors of computer arithmetic, we cannot call the analysis results 100\% reliable. These analysis discrepancies can have detrimental consequences in case of safety-critical applications like forest fire detection, e.g.,
a potential fire threat may be ignored due to an undetected system bug.

In order to overcome the common drawbacks of simulation, formal methods \cite{Gupta_92}
have been proposed as an efficient solution to validate a wide range of hardware and software systems. Formal methods enhance the analysis reliability by rigorously using mathematical
techniques to analyze the mathematical model for the given system.
The need for such mathematical methods in the context of WSNs is highlighted in \cite{McIver_06}.
However, the usage of  formal methods for probabilistic analysis is quite restricted.
The main limitation here is that the
random components of the system cannot be directly modeled using traditional formal tools.
For example, it is impossible to precisely reason about statistical
properties, such as expectation and variance, in the case of state-based approaches.
Furthermore, traditionally, huge proof efforts are expected to be involved in reasoning
about random components of a wireless system in the case of theorem proving.
Due to some recent developments in the higher-order-logic formalization of probability theory
\cite{Hurd_02,Hasan_08}, the analysis of a variety of wireless systems
with random components can be handled in a higher-order-logic theorem prover \cite{Gordon_89}
with reasonable amount of proof effort.

In this paper, we propose to formally verify, for the first time, the design of a real-world WSN
for forest fire detection using the k-set
randomized algorithm \cite{Liu_10} as an energy saving mechanism.
In particular, we perform a probabilistic
analysis using theorem proving to verify performance characteristics of a WSN deployed
for forest fire detection. In particular, we verify the coverage
properties related to the expected values of coverage intensity, the upper
bound on the total number of disjoint subsets, for a given coverage intensity,
and the lower bound on the total number of nodes.
This verification is based on our prior formalization of the k-set randomized algorithm, done in \cite{ICFEM_11}, and developed within the probabilistic framework \cite{Hasan_08} available in the HOL theorem prover.

The rest of this paper is organized as follows. First, we discuss some related work in Section 2.
Section 3 provides the formal probabilistic analysis
of a coverage-based random scheduling algorithm.
We utilize this foundational development to formally verify
a real-world WSN application for
forest fire detection in Section 4. We finally conclude the paper in Section 5.

\section{Related Work}

In \cite{forest_app}, the k-set randomized scheduling is used in order to save energy within a novel system for forest fire detection using Wireless Sensor Networks.
The performance evaluation is done by paper-and-pencil analysis and focuses on the detection accuracy of a forest fire. Experimentation is finally done to validate the forest fire system on a real prototype of 5 nodes.
A general surveillance framework composed of sensor nodes and robots is presented in \cite{Xiao_09}. Such framework is proposed to be used for environmental monitoring applications such as forest fire detection. Theoretical analysis is conducted to validate the performance of the network coverage.
Results are validated through simulation on a circular surface of a radius $R$ = 10000, where up to $n$ = 2000 nodes are uniformly deployed.
In \cite{Liu_10}, a coverage-based random scheduling algorithm
has been analyzed by a mathematical model. The coverage of this algorithm is
subsequently enhanced in \cite{Lin_08} by eliminating some blind points.
Both evaluations have been done using a Java simulator
by setting the monitored region to 200m$\times$200m, the detection range to 10m, and the number of sub-networks to 6.
Due to the inherent nature of simulation coupled with the usage of computer arithmetic,
these probabilistic analysis results cannot be called accurate.
Moreover, the analysis results are not generic, i.e., they are specific to a particular region area, range and number of sub-networks.

Probabilistic model checking
has been successfully used
for the probabilistic analysis of wireless systems \cite{Rutten_04}.
Probabilistic model checking has the same principles as traditional model checking:
the mathematical model of the probabilistic system is exhaustively
tested to check if it meets a set of probabilistic properties.
This technique has been successfully used to validate many aspects of WSNs
\cite{Zayani_10,Ballarini_06}. Moreover, the authors of \cite{OGDC_07} performed the formal analysis of the Optimal Geographical Density Control algorithm, called also OGDC,
in the RT-Maude rewriting tool \cite{RTurl}.
In addition to its accuracy, the main advantage of probabilistic model checking
methods is its mechanization.
However, it also suffers from some major shortcomings like the common problem of state-space
explosion \cite{Clarke_00}, and the inability to accurately reason about statistical properties.
Such problems have been noticed in \cite{Zayani_10,Ballarini_06,OGDC_07}

We overcome the limitations of both simulation and model checking techniques
by using the probabilistic framework developed in the HOL theorem prover
to formally verify a forest fire detection application deploying a WSN
using a variant of the randomized scheduling of nodes.
The HOL probabilistic framework is principally based on Hurd's PhD thesis \cite{Hurd_02} where the formalization of some discrete random variables along with their verification, based on the corresponding probability mass function (PMF) properties is presented.
In \cite{Hasan_08}, Hurd's formalization framework has
been extended with a formal definition of expectation. This definition is then utilized to
formalize and verify the expectation and variance characteristics associated with discrete
random variables that attain values in positive integers only.
Statistical properties of continuous random variables have been also verified in \cite{Hasan4_09}.
These foundations have been used to formally analyze various real-world applications including
the Miller-Rabin primality test; a well-known and commercially used probabilistic algorithm \cite{Hurd_03}, the stop-and-wait protocol \cite{Hasan2_09}, a stuck-at
fault model for the reconfigurable memory arrays \cite{Hasan1_09} and the automated repeat request (ARQ) mechanism at the logic link control (LLC) layer of the General
Packet Radio Service (GPRS) standard for Global System for Mobile Communications (GSM) \cite{Hasan3_09}. However, to the best of our knowledge, this is the first time that the probabilistic analysis using theorem proving technique is applied to analyze a forest fire detection WSN application in this paper.

\section{Formal Analysis of the k-set Randomized Scheduling Algorithm for WSNs}

In this section, we give an overview
of the formalization of the k-set randomized scheduling
algorithm for WSNs using the HOL theorem prover \cite{ICFEM_11}.
We will build upon these foundations to formally verify the forest fire detection WSN properties in the next section.

Consider a WSN that is formed by randomly deploying $n$ nodes over a two-dimensional field of interest.
Every sensor in this WSN can only sense the environment and detect events within its sensing range $r$.
During the initialization phase, the k-set randomized scheduling is run on each of the nodes as follows.
Each node starts by randomly picking a number ranging from $0$ to $(k-1)$. We denote the selected number by $i$.
Now, the node is assigned to the sub-network $i$ and will be turned on
only during the working time slot $T_i$ of that subset. During the other time slots,
it will be in the idle state.
Hence, during the time slot $T_i$, only the nodes belonging to the sub-network \emph{i}
will be active and can detect an occurring event.
The scheduling algorithm terminates by creating $k$ disjoint sub-networks that work independently and alternatively so that the energy over the whole network can be preserved.
It is important to note that each node joins a single subset with the same
probability $1/k$ since nodes are uniformly and independently deployed over the area of interest.

\begin{figure}[h]
\centering
\includegraphics[width=3in]{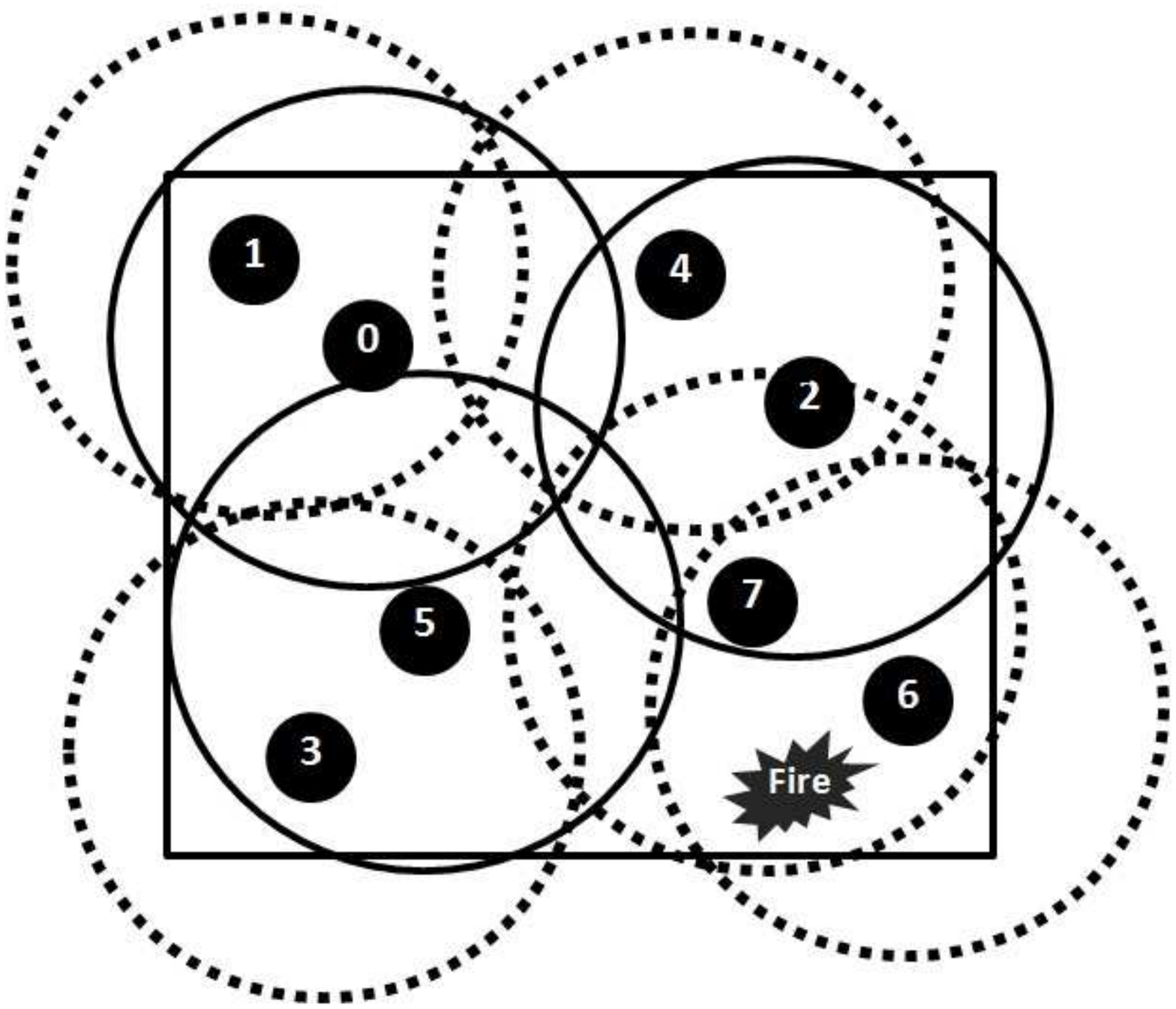}
\caption{An example of the k-set randomized scheduling for 8 nodes.}
\label{rd_sch_fig}
\end{figure}

For illustration purposes, Fig. \ref{rd_sch_fig} shows how the k-set randomized scheduling algorithm
splits arbitrarily a small WSN of eight sensor nodes to two sub-networks.
The eight nodes, randomly deployed in the monitored region, are identified by IDs ranging from 0 to 7.
The two sub-networks are called $S_0$ and $S_1$. Each node randomly chooses a number
0 or 1 in order to be assigned to one of these two sub-networks. Suppose that nodes 0; 2; 5 select the number 0 and join the subset $S_0$ and nodes 1; 3; 4; 6; 7 choose the number 1 and join the subset $S_1$. These two sub-networks will work alternatively, i.e., when the nodes 0; 2; 5, with sensing ranges denoted by the solid circles, are active, the nodes 1; 3; 4; 6; 7 illustrated by the dashed circles will be idle and vice-versa.

Since the assignment of the sensor nodes to the $k$ sub-networks is randomly done,
it may happen that some of the sub-networks are empty.
Moreover, due to the random deployment of nodes, the random scheduling
can lead to a situation where certain parts of the area are not
monitored at all or simultaneously monitored by many sensors.
While analyzing scheduling schemes, we are usually interested in
finding the probability that
a fire can be detected at each point of the forest by at least one active sensor.
Each point of the area is hence characterized by a coverage intensity $C_p$, which is defined as the average time during which the point is covered in a scheduling cycle \cite{Liu_10}.

\begin{equation}\label{Eq2_Cp}
C_p = \frac{{E[X] \times T}}{{k \times T}}
\end{equation}
\noindent where $E$[$X$] denotes the expectation of the random variable $X$
describing the total number of non-empty subsets. 

\begin{equation}\label{Sum_RV_Ber}
X = \sum\limits_{j = 0}^{k - 1} {X_j}
\end{equation}

\noindent where $X_j$ is the Bernoulli random variable whose value is $1$ in case of a non-empty subset. The coverage intensity of a WSN using k-set randomized scheduling has been formalized in higher-order logic as a function denoted by \texttt{cvrge\_intsty\_pt} \cite{Elleuch_11}. This function accepts two parameters, i.e., the number of sub-networks $k$ and the number of nodes $c$ covering a specific point inside a field. It utilizes the formalized Uniform and Bernoulli random variables to return the coverage intensity of a point using Equations (1) and (2).

Based on this formalization, the following mathematical expression for coverage intensity of a point has been formally verified in \cite{Elleuch_11}.

\begin{flushleft}
\texttt{\textbf{Theorem 1.}\\
$\vdash$ $\forall$ c,k. cvrge\_intsty\_pt c k = 1 - (1 - (1$/$(k+1)))$\mathtt{^c}$.}
\end{flushleft}

\noindent The proof details of the above theorem can be found in \cite{Elleuch_11}.

The variable \emph{c} above can be modeled by a Binomial random variable with success probability
$q$, i.e., the probability that a sensor covers a given point. Using this fact, the coverage intensity of the whole WSN with $n$ nodes has been formally defined in \cite{Elleuch_11} as the function \texttt{cvrge\_intsty\_network}.

\begin{flushleft}
\texttt{\textbf{Definition 1.}\\
$\vdash$ $\forall$ q,n,k. cvrge\_intsty\_network q n k = \\
\ \ expec\_fn ($\mathtt{\lambda}$x. 1 + -1$\times$(1 - 1$/$(k+1))$\mathtt{^x}$)
(prob\_binomial\_p n q).}
\end{flushleft}

\noindent The function \texttt{(prob\_binomial n q)} represents the Binomial random variable with $n$ trials and success probability $q$ \cite{Hasan_08} and the function \texttt{expec\_fn} represents the expectation of a function of a random variable \cite{Hasan_08}. While the functions of type
($\lambda x.\:C\:x$) represent the lambda abstraction functions in HOL that accept a parameter $x$ and return $C x$.

The following alternate mathematical expression for \texttt{cvrge\_intsty\_network} has also been formally verified in \cite{ICFEM_11}.

\begin{flushleft}
\texttt{\textbf{Theorem 2.}\\
$\vdash$ $\forall$ n,q,k. (0 $\mathtt{\leq}$ q) $\wedge$ (q $\mathtt{\leq}$ 1) $\wedge$ (1 $\mathtt{\leq}$ n) $\Rightarrow$ \\
\ \ \ \ \ \ \ \ \ \ \ (cvrge\_intsty\_network q n k = (1 - (1-(q$/$(k+1)))$\mathtt{^n}$)).}
\end{flushleft}

\noindent The assumptions of the above theorem ensure that the probability $q$ lies in the interval [0,1] and the number of nodes is at least 1.
In this paper, we build upon the above mentioned foundational results to analyze a WSN based forest fire detection application using the HOL theorem prover.

\section{Formal Analysis of WSN for Forest Fire Detection}

Consider a forest region of size $a$ = 100m$\times$100m, and
a WSN with sensors with sensing range $r$ = 30m \cite{Xiao_10}.
The main job of the sensors is to sense and communicate
temperature, humidity and barometric pressure values to a base station.
The k-set randomized scheduling is applied as an energy saving approach for this
forest fire detection application \cite{forest_app}.

\subsection {Formal Specification of the WSN for Forest Fire Application}

The first step in the formal probabilistic analysis using theorem proving for any application is to formally specify the system behavior using a higher-order-logic function \cite{Hasan_08}. We can formally specify the given forest fire detection application by specializing Definition 1 since it describes the generic coverage intensity of a WSN using a k-set randomized scheduling algorithm. In the given application, the success probability $q$ of a sensor covering a point is given by the ratio of the radius covered by a sensor with the total area, i.e., $q$ = $r$/$a$ = 30/10000 = 0.003. Thus, the coverage network intensity of the given forest fire detection using WSN can be formalized as follows:

\begin{flushleft}
\texttt{\textbf{Definition 2.}\\
$\vdash$ $\forall$ n,k. (1 $\mathtt{\leq}$ n) $\wedge$ (0 < k) $\Rightarrow$ \\
\ \ \  cvrge\_intsty\_forest\_WSN n k = cvrge\_intsty\_network 0.003 n k.}
\end{flushleft}

\noindent The above definition accepts two parameters, i.e., the total number of sensor nodes $n$ and
the number subsets $k$. It returns the coverage intensity of the system as the average value using Definition 1.

\subsection {Formal Verification of Probabilistic Properties}

The next step in the probabilistic analysis using the theorem proving approach is to specify the properties of interest as higher-order-logic proof goals and verify them in a theorem prover. For our given forest fire detection application, we verified the following theorem related to its coverage intensity:

\begin{flushleft}
\texttt{\textbf{Theorem 3.} \\ 
$\vdash$ $\forall$ n,k. (1 $\mathtt{\leq}$ n) $\Rightarrow$ \\
\ \ \  (cvrge\_intsty\_forest\_WSN n k = (1 - (k+0.997$/$(k+1))$\mathtt{^n}$)).}
\end{flushleft}


\noindent The proof of the above theorem in HOL is based on Theorems 1
and 2 and some arithmetic reasoning.
Theorem 3 plays a vital role in deciding the values for parameters $n$ and $k$ as demonstrated next.

\subsection {Formal Asymptotic Analysis of Probabilistic Properties}

Based on the formal verification of probabilistic coverage done so far,
we now conduct a formal asymptotic analysis of the basic WSN parameters
$n$ and $k$.
The number of deployed nodes $n$ is a common critical attribute which has
significant impact on both energy and coverage.
Deploying too few nodes may not guarantee a good coverage,
whereas deploying too many nodes can lead to a waste of energy.
Thus, an interesting design parameter is the minimum number of nodes $n\_min$ that are required
to deploy in order to ensure a network coverage intensity $C_n$ of at least $t$,
for a given number $k$.
This lower bound can be formally verified from Theorem 3 which gives a clear relationship between the network coverage
intensity $C_n$, the number of nodes $n$ and the number of disjoint sub-networks \emph{k}.
Now, in the case of the forest fire application, suppose that
a network coverage intensity of at least 70\% is targeted \cite{Xiao_10},
then the upper bound on the number of required nodes $n$ is verified in Theorem 4.
\begin{flushleft}
\texttt{\textbf{Theorem 4.} \\
$\vdash$ $\forall$ n,k. (1 $\mathtt{\leq}$ n) $\wedge$ (0 < k) $\wedge$ (t $\mathtt{\leq}$ cvrge\_intsty\_forest\_WSN n k) $\Rightarrow$\\
\ \ \ \ \ \ \ \ $\left[ {\frac{{\ln (1 - 0.7)}}{{\ln \left( {1 - \frac{0.003}{k}} \right)}}} \right] \le$ n.}
\end{flushleft}

\noindent The proof of the above theorem is based on Theorem 3, some properties of transcendental functions and arithmetic reasoning. It can be used to deduce useful results of the given application.
For example, we can deduce that under the randomized scheduling that divides the network into $k$ = 4 sub-networks,
at least $1606$ nodes are required to be deployed over the
forest area in order to achieve a network coverage intensity of 70\%.

After the nodes deployment, the number of nodes becomes known and fixed.
In the case of WSN applications which do not use the randomized scheduling,
the only way to enhance the network coverage intensity would be to deploy more nodes
so that the required coverage intensity can be achieved.
However, a second deployment can be very costly in the context
of such inhospitable field, when nodes are generally deployed by throwing them from an airplane.
Contrarily, for the WSN deployed for forest fire detection using the k-set randomized
scheduling, it is possible to increase the
coverage by updating the number of disjoint subsets \emph{k} by a suitable value.
We can formally deduce that for a given \emph{n} and a network coverage
intensity of at least \emph{t}, the upper bound on the number of disjoint subsets \emph{k} is
given as follows:

\begin{flushleft}
\texttt{\textbf{Theorem 5.} \\
$\vdash$ $\forall$ n,k. (1 $\mathtt{\leq}$ n) $\wedge$ (0 < k)
 $\wedge$ (0.9 $\mathtt{\leq}$ cvrge\_intsty\_forest\_WSN n k) \\
\ \ \ \ \ \ \ \ \ $\Rightarrow$
$k \le \dfrac{0.003}{{1 - {e^{{\textstyle{{\ln (1 - 0.9)} \over n}}}}}}.$}
\end{flushleft}

\noindent This result formally confirms the general intuition about the randomized scheduling approach.
Increasing $k$ saves energy, but leads to several sub-networks, which in turns translates to a poor network coverage intensity $C_n$.
This can decrease the performance of the whole network, which makes the accuracy in the probabilistic analysis of the value of $k$ very important after the deployment.
We also formally verified that in order to enhance the network coverage intensity from 70\% to 90\%
the value of $k$ has to be updated to 2, but it should not exceed 2 for the given values of $n$ and $t$.
Within the deployed WSN, the adjustment of $k$ is done by message flooding which informs all nodes about the new value.

Compared to classical techniques for analyzing a WSN for forest fire detection, using
the k-set randomized scheduling algorithm, our approach is much more efficient.
Indeed, while paper-and-pencil based analysis \cite{Xiao_09} or simulation \cite{forest_app}
cannot guarantee the correctness of the scheduling performance results,
the reported theorems in this paper are 100\% accurate.
This distinguishing feature is due to the inherent soundness of theorem proving and its generic nature, e.g., the coverage intensity for any given values of $n$ and $k$ can be computed by instantiating Theorem 3 with appropriate values.
Moreover, for each of the formally verified theorems,
the set of required assumptions is clearly stated so there is no doubt about missing a critical assumption. This feature is not ensured by classical analysis techniques where
many assumptions can be implicitly taken into account without a clear statement.

The above mentioned additional benefits are attained at the cost of the time and effort spent,
while conducting the HOL formalizations of the k-set randomized scheduling and its properties, by the user. The formal analysis in HOL took about
200 man-hours of extensive work.

\section{Conclusions}

Forest fire detection is one of the most safety-critical applications of WSN. Traditionally, the probabilistic analysis in this domain is conducted using paper-and-pencil proof methods or simulation. These techniques cannot ascertain 100\% accuracy and thus their usage may risk the reliability of fire detection.
In this paper, we propose to overcome this limitation by using formal methods. We built upon the previously verified results associated to the expected values of
the coverage intensity to provide, for the first time, a reliable probabilistic analysis of the design of a real-world WSN application for forest fire detection.
Our formally verified results include the expected values of coverage intensity, the upper bound on the total number of disjoint subsets, for a given expected
coverage intensity, and the lower bound on the total number of nodes.
These results are critical design parameters for any forest fire detection using WSN application and their accurate assessment would in turn ensure more reliable forest fire detection mechanisms. An interesting future direction related to the context of this paper is to analyze the detection delay in the forest fire detection application by building upon the foundations presented in \cite{ICFEM_11}.

\nocite{*}
\bibliographystyle{eptcs}
\bibliography{BiblioSCSS}

\end{document}